\documentstyle[epsfig,12pt]{article}
\newcommand{\pp}{{1\over \hat{\epsilon}}}
\newcommand{\p}{-{1\over \hat{\epsilon}}}
\newcommand{\ppme}{{{1-\epsilon}\over \hat{\epsilon}}}
\newcommand{\pe}{-{{1+\epsilon}\over \hat{\epsilon}}}
 \newcommand{\ppe}{{{1+\epsilon}\over \hat{\epsilon}}}
 \newcommand{\pme}{-{{1-\epsilon}\over \hat{\epsilon}}}
\begin{document}
\parskip 0.3cm
\begin{titlepage}
\begin{flushright}
CERN-TH/96-261\\
 hep-ph/9610362
\end{flushright}

\begin{centering}
{\large {\bf Next to Leading Order QCD Corrections  \\ to Polarized $\Lambda$
Production in  DIS}}\\
 \vspace{.4cm}
{\bf  D. de Florian\footnote{On leave of absence from Departamento de 
F\'\i sica, Universidad de Buenos Aires, Ciudad Universitaria Pab.1 (1428) Bs.As., Argentina. }}
\vspace{.05in}
\\ {\it Theoretical Physics Division, CERN, CH 1211 Geneva 23, Switzerland}\\
e-mail: Daniel.de.Florian@cern.ch \\
 \vspace{.4cm}
{\bf R. Sassot} \\
\vspace{.05in}
{\it  Departamento de F\'{\i}sica, 
Universidad de Buenos Aires \\ 
Ciudad Universitaria, Pab.1 
(1428) Bs.As., Argentina} \\
\vspace{1.5cm}
{ \bf Abstract}
\\ 
\bigskip
{\small
We calculate next to leading order QCD corrections to  semi-inclusive 
polarized deep inelastic scattering  and $e^+e^-$ annihilation cross sections 
for processes where the polarization of the identified final-state hadron can  
also be determined. Using dimensional regularization and the HVBM prescription 
for the $\gamma_5$ matrix, we compute corrections for different spin-dependent 
observables, both in the $\overline{MS}$ and $\overline{MS_p}$ factorization 
schemes, and analyse their structure. In addition to the well known corrections 
to polarized parton distributions, we also present those for final-state polarized 
fracture functions and polarized fragmentation functions, in a consistent 
factorization scheme.}  
\end{centering}
\vfill
\begin{flushleft}
CERN-TH/96-261\\
October 1996 \hfill
\end{flushleft}
\end{titlepage}
\vfill\eject

\noindent{\large \bf 1. Introduction}\\

In recent years, a considerable degree of attention has been paid to the 
semi-inclusive production of  $\Lambda$  hyperons and their polarization.
The approaches include experimental programmes to measure polarized fragmentation 
functions in $e^+e^-\rightarrow \Lambda+X$ \cite{burjaf} and semi-inclusive  
lepton-proton deep inelastic scattering (DIS) \cite{comp,jaf}, both in the 
current fragmentation region, as well as models for the production of these 
hyperons in existent $p\bar{p}$ and DIS experiments, but in the target 
fragmentation region \cite{elko}. Many of these studies are closely related 
to the interpretation of the proton spin structure \cite{elliot,elka}, due to 
the possibility of reconstructing it from the observed $\Lambda$ polarization.

Notwithstanding this increasing experimental interest in spin-dependent 
semi-inclusive  $\Lambda$ physics, and also the important phenomenological 
insights related to the production of these hyperons in different processes,
QCD corrections to such cross sections have received little attention.
These corrections, however, have been shown to be crucial for the interpretation
of totally inclusive polarized DIS experiments related to the spin
structure of the proton \cite{elliot,elka}, and they certainly would have a  
non-negligible role  in polarized $\Lambda$ production. On the other hand, one 
would naturally expect any parton  model-inspired phenomenological description 
of the processes to mould into the more formal QCD-improved description, at 
least in some adequate limit. A description including fracture functions \cite{ven,graudenz}
 allows not only a consistent factorization of the collinear divergences 
characteristic of semi-inclusive processes, but also  a perturbative 
picture of events where the identified  final-state hadron comes from the
target fragmentation region.
 
In addition to their phenomenological interest, QCD corrections to polarized 
semi-inclusive processes involve some theoretical subtleties, such as factorization 
procedures \cite{vogel}, which in this case have to be extended to polarized 
fragmentation functions, and final-state polarized fracture functions, 
and which deserve a very careful treatment.
In a recent paper \cite{npb} we have computed the $\cal{O}$$(\alpha_s)$ corrections 
to the one-particle inclusive polarized DIS cross sections, 
using dimensional regularization \cite{bollini} and the HVBM \cite{HVBM} 
prescription for the $\gamma_5$ matrix, discussing different factorization 
schemes. The analysis includes also target fragmentation effects by means of 
polarized fracture functions,  but   the polarization of the identified 
hadron is not taken into account, given that the observables  under study are 
those related to pions and kaons in the final state.

In this paper we extend the approach  of ref. \cite{npb} in order to 
include the possibility of polarized hadrons, such as the $\Lambda$, in the 
final state. This inclusion requires the computation of $\cal{O}$$(\alpha_s)$ 
corrections to polarized fragmentation functions, which dominate $\Lambda$ 
production in the current fragmentation region and  can be measured in 
$e^+e^- \rightarrow \Lambda + X$, and also to final-state polarized fracture 
functions, which, as it has been said, account for target fragmentation 
effects and make possible a consistent factorization of collinear 
singularities in semi-inclusive DIS.

Next to leading order (NLO) corrections to polarized fragmentation functions 
have been addressed recently in refs. \cite{rav}, using massive gluons 
to regulate infrared singularities. In this paper, however, we keep dimensional 
regularization and we propose a factorization prescription similar to  the ones 
of our previous work \cite{npb,prd}. The factorization prescriptions for 
final-state polarized fracture functions can be generalized directly from those for 
polarized parton distributions, which guarantee the usual current conservation 
properties \cite{gordon,prd}.
The use of a common regularization procedure and natural factorization 
prescription allows a consistent and  more direct  analysis of cross sections 
for different processes. 
 
In the following section we introduce the formal framework for obtaining the
QCD corrections to polarized fragmentation functions in $e^+ e^-$ annihilation
processes, and we specify our choice for factorization. In the third section 
we do the same for fracture functions, discussing there  the structure of the 
different cross sections  resulting from the combinations of polarization in 
the initial and final states. Finally, we summarize our results and present 
conclusions.\\

\noindent{\large \bf 2. $e^+ e^-$ annihilation}\\

In this section we obtain the NLO corrections for the semi-inclusive production
of  a longitudinally polarized hadron ($h$) in $e^+ e^-$ annihilation. This 
process, besides  offering the clearest way to define at leading order, and 
eventually measure, the polarized fragmentation functions, allows us to show  
our choice for the factorization prescription, which defines the fragmentation 
functions beyond the leading order.

These corrections have recently been obtained in ref. \cite{rav}, using off-shell gluons as regularization prescription. Since we want to make contact 
with previous calculations for other semi-inclusive DIS processes,  which were 
performed  using dimensional regularization,  we first obtain the above-mentioned 
corrections using the latter prescription.

The usual kinematical variables used for the description of processes like 
the one under consideration, i.e. $e^+(l) e^-(l') \rightarrow h(p)+ X$, are
\begin{equation}
z=\frac{2\, p.q}{Q^2} \,\,\,\,\, {\rm and} \,\,\,\,\, Q^2=q^2 ,
\end{equation}
where the variable $q$ denotes the four-momentum of the virtual photon, and
is defined by the momentum of the incoming leptons, $q=l+l'$.
In terms of these variables it is  possible to write down the photon 
fragmentation tensor $W_{\mu\nu}$, whose antisymmetric part $W_{\mu\nu}^A$ is
\begin{equation}
W_{\mu\nu}^{A} = 2 \frac{i N_c}{z^2 Q^2} \epsilon_{\mu\nu\rho\sigma} q^\rho s^\sigma g^h_1 .
\end{equation}
We do not include  here the piece related to the additional semi-inclusive structure
function $g^h_2$, which only contributes in the case of transverse polarization.

Taking the difference between the cross sections for the processes where the 
incoming electron and the outgoing hadron are polarized parallel and antiparallel 
to each other, we obtain
\begin{equation}
\frac{d\sigma (\uparrow\uparrow- \downarrow\uparrow)}{dz\, d(\cos\theta)} = 
\alpha^2 \frac{N_c \pi}{Q^2} g^h_1 \cos\theta ;
\end{equation}
 here $\theta$ is the angle between the hadron and the beam directions.

In the naive parton model, the polarized structure function $g^h_1 (z)$  is 
given by the convolution between the partonic cross section corresponding
to the Feynman diagram of fig. 1a and the polarized semi-inclusive 
fragmentation functions
\begin{equation} 
\Delta D_{h/q}(z)= D^{h\uparrow}_{q\uparrow} (z) - D^{h\uparrow}_{q\downarrow} (z) ,
\end{equation}
i.e.
\begin{equation}
\label{naive}
g^h_1 (z)= \sum^{N_f}_q e_q^2  [\Delta D_{h/q}(z)+\Delta D_{h/{\bar{q}}}(z)] .
\end{equation}

In NLO of QCD, the interference between the diagrams of figs. 1a and 1b,  that between the first one and those in 1c and 1d, 
as well as the whole contributions from the diagrams of fig. 2 are included, 
in both cases taking into account that either quarks or gluons undergo 
hadronization. Then, eq. (\ref{naive}) develops the less trivial  
convoluted expression:
\begin{eqnarray}
\label{1loop}
g^h_1 (z,Q^2) = \sum^{N_f}_q e_q^2   \int_z^1 \frac{dy}{y} \left\{\left[ \delta(1-y)+ 
\Delta C_q \left(y, \frac{Q^2}{\mu^2}\right)\right] \times \right. \nonumber \\ 
\left. \left[\Delta D_{h/q}\left(\frac{z}{y}\right) +  
\Delta D_{h/{\bar{q}}}\left(\frac{z}{y}\right) \right] 
 +\Delta  C_g\left(y, \frac{Q^2}{\mu^2}\right) \Delta D_{h/g}\left(\frac{z}{y}\right) \right\} .
\end{eqnarray}

Since in some intermediate steps of the calculation, both infrared and soft 
collinear divergences appear, (the UV divergences can be avoided by working in the 
Landau gauge), it is necessary to define a prescription to isolate them. For 
this purpose,  and throughout this paper, we use dimensional regularization  
and the HVBM scheme, which  allows us to deal with  typically four-dimensional 
objects such as $\epsilon^{\mu\nu\rho\sigma}$ and $\gamma^5$ in a fully 
consistent way. As  is well known \cite{vogel,webber},  
spurious terms appear in this scheme because of the breaking of the chiral symmetry. These terms 
can be absorbed in the factorization procedure in order to keep the naive 
interpretation of the first moment of the polarized fragmentation function, 
i.e. as the fraction of the spin of the parton carried by the outgoing hadron.

In order to calculate the NLO corrections, it is useful to define the  variables 
$x_i=2\, k_i . q/ Q^2 = 2\, E_i/Q$, in terms of the momentum carried by the three-final state partons. For the sake of simplicity, we choose the $z$ direction as 
that of the parton labelled by the index 1, and  $\phi$ as the angle between 
the parton labelled  2 and the $z$ direction.  Momentum conservation implies that
\begin{equation}
 x_2 = \frac{1-x_1}{1-x_1(1-y)},
\end{equation} 
where 
\begin{equation}
y=\frac{1+\cos\phi}{2},
\end{equation}
and integrating over the particles labelled  2 and 3
(particle 1, which undergoes hadronisation, is not integrated), the three ($2+1$) particle 
phase space reduces to
\begin{equation}
PS^{(2+1)} = \frac{(4\pi)^\epsilon}{8\pi}\int_0^1 dy \, 
\frac{x_1^{1-2\epsilon}}{1-x_1}\, x_2^2\, \int_0^{Q^2 x_2^2 y(1-y)} 
d|\hat{k}|^2 \, \frac{|\hat{k}|^{-2(1+\epsilon)}}{\Gamma (-\epsilon)}.
\end{equation}
Here $|\hat{k}|^2$ is the square of the $d-4=-2 \,\epsilon$ dimensional 
component of the momentum of the outgoing (integrated) particles 2 and 3.

Projecting conveniently the matrix elements corresponding to the diagrams in 
fig. 2,  and integrating over the phase space, we obtain the real emission 
contributions to $\Delta C_q$ (we label the outgoing quark as number 1):
\begin{eqnarray}
\label{realq}
\Delta C_q \left(y,\frac{Q^2}{\mu^2}\right) &=& 
\left( \frac{4\pi
\mu^2}{Q^2} \right)^\epsilon \frac{\Gamma(1-\epsilon)}{\Gamma(1-2 \epsilon)}
\frac{\alpha_s}{2\pi} C_F \left\{ -\frac{1}{\epsilon} \left[ \frac{3}{2} 
\delta(1-y) + \frac{1+y^2}{(1-y)_+} \right] \right.\nonumber \\
  &+& \left.  \, \delta(1-y) \left[ \frac{2}{\epsilon^2} + \frac{3}{\epsilon} 
+ 8 -\frac{2 }{3} \pi^2 \right] + \Delta f_q^D(y) \frac{}{} \right\}, \,\,\,\,\,\,\,\,
\end{eqnarray}
with
\begin{eqnarray}
\Delta f_q^D (y) &=&  \delta(1-y) \left[ \frac{2}{3} \pi^2 -\frac{9}{2} 
\right] -\frac{3}{2} \left(\frac{1}{1-y}\right)_+ + (1+y^2) 
\left(\frac{\ln (1-y)}{1-y}\right)_+  \nonumber \\
 &+& 2\, \frac{1+y^2}{1-y} \ln y -\frac{3}{2} (1-y) -2 \widehat {(1-y)} . 
\end{eqnarray}
In the last equation, we have kept track of terms originated in the 
$4-2\, \epsilon$ dimensional momentum integration, writing them under  hats just 
for factorization purposes.
The second term in the r.h.s. of eq. (\ref{realq}), which contains the 
infrared divergences, cancels identically when adding the virtual contributions 
coming from the interference of diagrams in fig. 1, as  is generally expected.

 The same procedure can be applied in order to obtain the gluonic coefficient
(now, labelling the gluon as parton 1) obtaining
\begin{eqnarray}
\label{realg}
\Delta C_g \left(y,\frac{Q^2}{\mu^2}\right) =  
\left(\frac{4\pi\mu^2}{Q^2} \right)^\epsilon \frac{ \Gamma(1-\epsilon)}
{\Gamma(1-2 \epsilon)}  \frac{\alpha_s}{2\pi} C_F \left\{ -\frac{1}{\epsilon} 
\left[ 2 \, (2-y) \right] + \Delta f_g^D(y) \right\} , 
\end{eqnarray} 
 with
\begin{eqnarray}
\Delta f_g^D (y) = 2\, \left\{ (2-y) \ln [(1-y)y^2]- (2-y) - 2 \widehat{(1-y)}
\right\} .
\end{eqnarray}

Having computed the whole cross section up to next to leading order, we are now 
able to factorize  the divergences by means of the definition of 
scale-dependent polarized fragmentation functions ($\Delta D_{h/q}(z,Q^2)$ and $\Delta D_{h/g}(z,Q^2)$). 
Fixing the factorization scale $\mu^2$ equal to $Q^2$, in the $\overline{MS}$ 
scheme the prescription amounts to absorbing only the $1/\hat\epsilon$ terms. It has 
been shown that within the HVBM method this prescription leaves unsubtracted some soft finite 
contributions directly  related to the hat terms. It is also 
frequent to define a different scheme, called $\overline{MS_p}$ \cite{gordon}, where these 
contributions are  subtracted. The general expression for the distributions,
whithout specifying any scheme, is then given by 
\begin{eqnarray}
\label{dd}
\Delta D_{h/q}(z) = \int^1_z  \frac{dy}{y} \left\{ \left[ \delta(1-y) +
\frac{\alpha_s}{2\pi} \left( \pp \Delta P_{q \leftarrow q} (y) - C_F \Delta 
\tilde{f}_q^D (y) \right) \right]  \times \right.  \,\,\,\, \nonumber \\
  \left.  \Delta D_{h/q}\left(\frac{z}{y},Q^2\right) + \left( \pp \Delta P_{g \leftarrow q}
  (y) - C_F \Delta \tilde{f}_g^D (y) \right) \Delta D_{h/g}\left(\frac{z}{y},Q^2\right) 
  \right\} , \,\,\,\,\,\,\,
\end{eqnarray}
where 
\begin{equation}
\pp \equiv {1\over \epsilon}
{{\Gamma[1-\epsilon]}\over{\Gamma[1-2
\epsilon]}}\left({{4\pi\mu^2}\over{Q^2}}\right)^\epsilon 
=\frac{1}{\epsilon}-\gamma_E+\log 4\pi+\log \frac{\mu^2}{Q^2}+\cal{O}(\epsilon)
\end{equation}

In the $\overline{MS_p}$ scheme, the finite subtraction terms $\Delta 
\tilde{f}_q^D (y) = \Delta \tilde{f}_g^D (y) = 4\, (y-1)$ are designed to absorb 
the soft contributions coming from the real gluon emission diagrams in fig. 2 
and, therefore, they are the same as the one obtained for the definition of NLO 
polarized quark distributions in DIS. In fact, the hat terms obtained for the 
real gluon emission diagrams in DIS are equal to the ones obtained in this 
analysis. 

Using the former definition for NLO distributions, the final expression for 
the semi-inclusive structure function reads
\begin{eqnarray}
\label{1loopfact}
g^h_1 (z,Q^2) = \sum^{N_f}_q e_q^2   \int_z^1 \frac{dy}{y} \left\{\left[ 
\delta (1-y) + \frac{\alpha_s}{2\pi} C_F \left(\Delta f_q^D(y) - 
\Delta \tilde{f}_q^D (y)\right) \right] \times \right. \nonumber \\
\left.
\left[\Delta D_{h/q}\left(\frac{z}{y}\right) +  
\Delta D_{h/{\bar{q}}}\left(\frac{z}{y}\right) \right] 
 + \frac{\alpha_s}{2\pi} C_F \left[ \Delta  f_g^D(y) - 2 \Delta  \tilde{f}_g^D(y)
\right]\Delta D_{h/g}\left(\frac{z}{y}\right) \right\} \,\,\,\,\,\,\,\,\,
\end{eqnarray}
where, as usual, in the $\overline{MS}$ scheme the finite subtracted terms 
are chosen to be zero.

Notice that although the full expression for the cross sections, and the
definition of scale-dependent densities, differs from those of refs.\cite{rav} in the finite terms, the evolution kernels, which 
are not scheme-dependent, are identical, as expected.\\

\noindent{\large \bf 3. Semi-Inclusive Deep Inelastic Scattering}\\

With the definition for polarized fragmentation functions given in the previous 
section, eq. (14), we are now able to compute the NLO corrections for 
semi-inclusive DIS in the case in which the final-state hadron is polarized.
NLO contributions for processes with unpolarized final-state hadrons (with 
either polarized or unpolarized initial states) have been computed in refs.
\cite{npb} and \cite{graudenz}, respectively, so  we refer the reader to these 
for most of the definitions and conventions. 
 
Using the usual kinematical DIS variables for the interaction between a 
lepton of momentum $l$ and helicity $\lambda_l$ and a nucleon $A$ of momentum 
$P$ and helicity $\lambda_A$
\begin{equation}
x=\frac{Q^2}{2\, P\cdot q} ,\,\,\,\,\,\,\, y=\frac{P\cdot q}{P\cdot l},\,\,\,\,\,\,\, Q^2=-q^2\,\, {\rm and}\,\,\,\, S_H=(P+l)^2 ,
\end{equation}
the differential cross section for the production of a hadron $h$ with energy $E_h= z\, E_A (1-x)$ and helicity $\lambda_h$ (with $n$ partons in the final state) can be written as
\begin{eqnarray}
\label{sec}
\frac{d \sigma^{\lambda_l \lambda_A \lambda_h}}{dx\,dy\,dz} &=&  \int \frac{du}{u}\sum_n\sum_{j=q,\bar q , g} \int dPS^{(n)}
\,  \frac{\alpha^2}{S_{H}x}\, \frac{1}{e^2 (2\pi)^{2d}}\,  \nonumber \\
&\times& \left[ 
Y_M(-g^{\mu\nu})+Y_L \frac{4x^2}{Q^2}P_\mu P_\nu + \lambda_l  Y_{P}\,
\frac{x}{Q^2}\, i \epsilon^{\mu \nu q P } \right]  \nonumber \\
&\times& \sum_{\lambda_1,\lambda_2 =\pm 1}H_{\mu\nu}(\lambda_1,\lambda_2)
\left\{  M_{j,h/A} \left( \frac{x}{u},\frac{E_h}{E_A},\frac{\lambda_1}{\lambda_A},\frac{\lambda_h}{\lambda_A}\right)(1-x) \right. \nonumber \\
&+& \left. f_{j/A}\left(\frac{x}{u},\frac{\lambda_1}{\lambda_A}\right)
\sum_{i_\alpha =q,\bar q , g }  D_{h/i_{\alpha}}\left(\frac{z}{\rho},
\frac{\lambda_h}{\lambda_2}\right)
\frac{1}{\rho} \right\}
\end{eqnarray}
where $\rho= E_\alpha/E_A$. The helicity-dependent partonic tensor is defined by \begin{equation}
H_{\mu\nu} (\lambda_1,\lambda_2) = M_{\mu}(\lambda_1,\lambda_2) 
M_{\nu}^\dagger(\lambda_1,\lambda_2) ,
\end{equation} 
where $M_\mu$ is the parton-photon matrix element with the photon polarization 
vector subtracted. Here $\lambda_1$ and $\lambda_2$ are the helicities of the 
initial- and final-state partons respectively. Both $f_{j/A}$ and $D_{h/i_{\alpha}}$ 
are the usual parton distribution and fragmentation functions, respectively, 
which represent the probabilities of finding a parton $j$ or a hadron $h$ with a given 
momentum fraction and helicity with respect from the parent particle (the target $A$ or the parton $i_\alpha$, respectively). 

As  was shown in ref. \cite{ven}, it is 
necessary to introduce a new   distribution  for semi-inclusive processes, the fracture function, which gives 
the probability of finding a polarized parton $j$ and a polarized hadron $h$ 
in the nucleon, denoted by $M_{j,h/A} \left( \frac{x}{u},\frac{E_h}{E_A},
\frac{\lambda_1}{\lambda_A},\frac{\lambda_h}{\lambda_A}\right)  $. 
Including them, it will be possible to factorize all the collinear 
divergences that remain in the NLO contributions after the redefinition of the 
scale-dependent parton distribution and fragmentation functions. 

Taking different linear combinations of cross sections for targets and
final-state hadrons with equal or opposite helicities, it is possible to  
isolate parton distributions, fragmentation functions and fracture functions
of different kind and their QCD corrections. For example, taking the sum 
over all the helicity states
\begin{equation}
\sigma=\sigma^{\lambda_l ++} +\sigma^{\lambda_l+-} +\sigma^{\lambda_l-+} 
+\sigma^{\lambda_l--},
\end{equation}
the result is proportional to a convolution of unpolarized parton distributions,
fragmentation  and fracture functions \cite{graudenz} \footnote{Notice the slight difference in the normalization used here (a factor of 2) with respect to ref. \cite{graudenz}, where an average over the polarizations of the target was taken.}. 
Taking the difference between cross sections with opposite target helicities, 
as in ref. \cite{npb}:
\begin{equation}
\Delta \sigma=(\sigma^{\lambda_l ++} +\sigma^{\lambda_l+-}) -(\sigma^{\lambda_l-+} 
+\sigma^{\lambda_l--}),
\end{equation}
the result contains polarized parton distributions, single polarized fracture 
functions (single initial-state polarization), and unpolarized fragmentation functions.
In the  following two subsections, we consider the two remaining possibilities,
which represent the case where only the final-state hadron polarization is put
in evidence (single final-state polarization) at cross section level:
\begin{equation}
 \sigma\Delta=(\sigma^{\lambda_l ++} -\sigma^{\lambda_l+-}) +(\sigma^{\lambda_l-+} 
-\sigma^{\lambda_l--}),
\end{equation}
and that where both polarizations are relevant (double polarization):
\begin{equation}
\Delta \sigma\Delta=(\sigma^{\lambda_l ++} -\sigma^{\lambda_l+-}) -(\sigma^{\lambda_l-+} 
-\sigma^{\lambda_l--}).
\end{equation}

\noindent{\large \bf 3a. Single (final-state) polarization}\\

It is straightforward to show that the single (final-state) polarization
cross section $\sigma \Delta$ is proportional to the convolutions 
\begin{eqnarray}
\label{conv}
\sigma \Delta \, \propto \, \, f_{j/A}(x/u)\otimes  H\Delta(u,\rho) \otimes
\Delta D_{h/i_{\alpha}}(z/\rho) \nonumber \\  
+(1-x) \, M\Delta_{j,h/A} (x/u, (1-x) z) \otimes 
\Delta H'(u)
\end{eqnarray}
where the first term corresponds to current fragmentation processes and
the second to fragmentation of the target.
In the former, $f_{j/A}(x/u)$ are just the unpolarized parton distributions, 
$\Delta D_{h/i_{\alpha}}(z/\rho)$ are the polarized fragmentation functions 
defined in section 2, and 
$H\Delta = \sum_{\lambda_1,\lambda_2} \lambda_2 \, H(\lambda_1,\lambda_2)$ is
the partonic tensor for polarized final states. The convolutions are those 
defined by eq.  (\ref{sec}). In the second term of eq. (\ref{conv}) we have defined the final state polarized fracture function as 
\begin{equation}
 M\Delta   = M_{j,h/A} \left( +,+\right)+
M_{j,h/A} \left( -,-\right) - M_{j,h/A} \left( +,-\right) - M_{j,h/A} \left( -,+\right),
\end{equation}
where we have omitted the  first two kinematical arguments in the fracture functions,
and the single polarized tensor $\Delta H'= \sum_{\lambda_1,\lambda_2} \lambda_1 \,H(\lambda_1,\lambda_2)$
(integrated in $\rho$).

Computing the matrix elements of the diagrams in figs. 4, 5 and 6, 
contracting with the adequate projector
\begin{eqnarray}
P_{pol}^{\mu\nu}\equiv  \frac{\alpha^2}{S_{H}x}\,\, \frac{1}{e^2 (2\pi)^{2d}}\,\, \frac{x}{Q^2} \,i \epsilon^{\mu \nu q P}, 
\end{eqnarray}
and integrating them over the phase space as it   done in ref.  \cite{npb}, we finally find:
\begin{eqnarray}
\label{cucusa}
&&\frac{d \sigma \Delta }{dx\,dy\,dz}  = 
  Y^p \lambda_l \sum_{i=q,\bar q} c_i \left\{ \int \int_{A} \frac{du}{u} \frac{d\rho}{\rho}
\left\{ q_i\left(\frac{x}{u}\right) \, \Delta D_{q_i}\left(\frac{z}{\rho}\right)\,
\delta(1-u)\delta(1-\rho)\frac{}{}  \right.\right.  \nonumber \\
&+& \left.\left.  q_i\left(\frac{x}{u}\right)\, \Delta D_{q_i}\left(\frac{z}{\rho}\right)\,
\right. \right. \nonumber \\
 &&\hspace*{-4mm}\times \left. \left. 
\frac{\alpha_s}{2\pi} \left[ \p \left(\Delta P_{q\leftarrow q}(\rho)\delta(1-u)+
P_{q\leftarrow q}(u)\delta(1-\rho)\right) +C_f 
\Phi\Delta_{qq}(u,\rho)\right]
 \right.\right. \nonumber\\
&+& \left.   \left.  q_i\left(\frac{x}{u}\right)\,\Delta D_g\left(\frac{z}{\rho}\right)\,
\right. \right. \nonumber \\
&&\hspace*{-4mm}\times \left. \left.
\frac{\alpha_s}{2\pi} \left[ \p  \left(\Delta P_{g\leftarrow q}(\rho)\delta(1-u)+
 \hat P\Delta_{gq\leftarrow q}(u)\delta(\rho-a)\right) +C_f 
\Phi\Delta_{qg}(u,\rho)\right] 
 \right. \right.\nonumber\\
&+ & \left.  \left.   g\left(\frac{x}{u}\right)\,\Delta D_{q_i}\left(\frac{z}{\rho}\right)\,
\right. \right. \nonumber \\
&&\hspace*{-4mm}\times \left. \left.
\frac{\alpha_s}{2\pi} \left[ \p  \left( P_{q\leftarrow g}(u)\delta(1-\rho)+
 \hat P\Delta_{q\bar{q}\leftarrow g}(u)\delta(\rho-a)\right) + T_f 
\Phi\Delta_{gq}(u,\rho)\right] \right\}\right.  \nonumber \\
& & \hspace*{40mm}\left. + \, \int_{B} \frac{du}{u} (1-x)
\left\{  M\Delta_{q_i}  \left(\frac{x}{u},(1-x) z\right)  \delta(1-u) \right. \right. \nonumber \\
&+&\left.\left.   M\Delta_{q_i}  \left(\frac{x}{u},(1-x) z\right)  \, \frac{\alpha_s}{2\pi} \left[ \p  \Delta P_{q\leftarrow 
q}(u)+C_f  \Phi\Delta_{q}(u) \right]  \right. \right.\nonumber \\ 
&+& \left.\left.  M\Delta_g \left(\frac{x}{u},(1-x) z\right) \,
 \frac{\alpha_s}{2\pi} \left[ \p \Delta P_{q\leftarrow g}(u)+ T_f 
\Phi\Delta_g (u)\right] \right\}\right\} 
\end{eqnarray} 
where the coefficients $\Phi\Delta_{ij}$ are given in  the Appendix,
\begin{equation}
c_j= 4\pi  Q^2_{q_j} \, {{\alpha^2}\over {S_H x}} \, ,
\end{equation}
 and the 
integration limits can be found in ref. \cite{npb}. 

The poles proportional to $\delta (1-u)$ and to the polarized Altarelli-Parisi  
kernels $\Delta P_{i\leftarrow j}$ \cite{ap} correspond to final-state singularities 
and are subtracted with the definition of the NLO polarized fragmentation functions 
given in  section 2. 
The coefficients $\Phi\Delta_{ij}$ contain finite soft terms (proportional to the terms marked with hats) related to polarized final-states which are also
 subtracted by means 
of eq.  (\ref{dd}) in the $\overline{MS_p}$ scheme.

Those poles proportional to  $\delta (1-\rho)$ and to the unpolarized kernels are 
related to the initial-state singularities and are therefore  subtracted by 
the NLO definition of unpolarized parton distributions. In this case both the
$\overline{MS}$ and the $\overline{MS_p}$ schemes coincide, since we are 
dealing here with  unpolarized initial-state processes, so there are no finite soft
terms.  Notice that there are no hat terms proportional to $\delta (1-\rho)$ 
in the coefficients.

Finally, the poles proportional to $\delta (\rho -a)$,  where $a=x(1-u)/u(1-x)$, represent configurations where the hadrons are produced by the current but in 
the direction of the incoming nucleon.  These are subtracted along with the 
target-initiated singularities  (homogeneous) by the NLO definition of 
final-state polarized fracture functions 
\begin{eqnarray}
\label{1p}
  M\Delta_{q_i} (\xi,\zeta) &=& \nonumber \\
  &&\hspace*{-25mm}\int_{\frac{\xi}{1-\zeta}}^1 \frac{du}{u} \left\{ 
  \left[ \delta(1-u) + \frac{\alpha_s}{2\pi} \left(
\pp \Delta P_{q\leftarrow q}(u) -  C_{f} \Delta 
\tilde{f}_q^{MH}(u)\right) \right]
 M\Delta_{q_i}\left(\frac{\xi}{u},\zeta,Q^2\right) \right. \nonumber \\
&  & \left. +\,\frac{\alpha_s}{2\pi} \left[
\pp \Delta P_{q\leftarrow g}(u) -  T_{f} \Delta 
\tilde{f}_g^{MH}(u) \right]   M\Delta_g\left(\frac{\xi}{u},\zeta,Q^2\right)
 \right\}\nonumber \\
& &\hspace*{-15mm} +\,\int_{\xi}^{\frac{\xi}{\xi+\zeta}} \frac{du}{u} \frac{u}{x(1-u)}
\left\{   \frac{\alpha_s}{2\pi} \left[
\pp  \hat P\Delta_{gq\leftarrow q}(u) - C_{f} \Delta 
\tilde{f}_q^{MI}(u) \right]
\right. \nonumber \\&& \left.  \hspace*{3mm} \times 
q_i\left(\frac{\xi}{u},Q^2\right)  
\Delta D_g\left(\frac{\zeta u}{\xi (1-u)},Q^2\right) \right. \nonumber \\
&&\left. + \, \frac{\alpha_s}{2\pi} \left[
\pp \hat P\Delta_{q\bar{q}\leftarrow g}(u) - \frac{\alpha_s}{2\pi} T_{f} \Delta 
\tilde{f}_g^{MI}(u) \right] 
\right. \nonumber \\ &&\left.  \hspace*{3mm} \times \, 
 g\left(\frac{\xi}{u},Q^2\right) \,
\Delta D_{q_i}\left(\frac{\zeta u}{\xi (1-u)},Q^2\right) \right\} \hspace*{10mm}
\end{eqnarray}
where the  homogeneous $(MH)$ and inhomogeneous $(MI)$ counterterms  in the 
$\overline{MS_p}$ are given by
\begin{eqnarray}
 \Delta \tilde{f}_q^{MI}(u) &=&  \Delta\tilde{f}_q^{MH}(u) = 4\,(u-1) \\ \nonumber
 \Delta \tilde{f}_g^{MI}(u) &=& \Delta\tilde{f}_g^{MH}(u)  = 2\,(1-u).
 \end{eqnarray}
These are identical to those in the final-state polarized case, since the 
structure of the corrections is the same
 (actually, the homogeneous coefficients are exactly the same).

One interesting feature of the divergences is that the poles corresponding 
to $\delta (1-\rho)$ and $\delta (\rho-a)$ in eq. (\ref{cucusa}) differ only in
a global sign $(-\hat P\Delta_{q\bar q \leftarrow g} (x) =  P_{q\leftarrow g} (x))$. 
This sign accounts for the opposite polarizations of the quark in the 
$\gamma g \rightarrow q\bar q$ process according to whether the quark is emitted in the 
photon direction ($\rho=1$) or in the opposite configuration ($\rho=a$).

Applying the factorization procedure, we obtain the following  cross section
 \begin{eqnarray}
\frac{d\sigma\Delta}{dx\,dy\,dz} & =& \nonumber \\
&&  \hspace*{-8mm} Y^p \lambda_l \sum_{i=q,\bar q} c_i\left\{\int \int_{A} \frac{du}{u}
\frac{d\rho}{\rho} 
\left\{ 
 q_i\left(\frac{x}{u},Q^2\right)\, \Delta D_{q_i}\left(\frac{z}{\rho},Q^2\right) \,
\delta(1-u)\delta(1-\rho)\frac{}{} 
\right. \right.  \nonumber \\
&+& \hspace*{-3mm}  \left.\left.
q_i\left(\frac{x}{u},Q^2\right)\, \Delta D_{q_i}\left(\frac{z}{\rho},Q^2\right)  \,
\frac{\alpha_s}{2\pi} C_f \left[ 
   \Phi\Delta_{qq}(u,\rho) - \Delta \tilde{f}_q^D(\rho) \delta(1-u) \right] \right.\right.
\nonumber \\   
&+& \hspace*{-3mm} \left.\left.  q_i\left(\frac{x}{u},Q^2\right)\, \Delta D_g\left(\frac{z}{\rho},Q^2\right)\,
\right.\right. \nonumber \\   
&& \hspace*{-3mm}\left. \left. \times \,\frac{\alpha_s}{2\pi} C_f \left[
 \Phi\Delta_{qg}(u,\rho)- \Delta \tilde{f}_g^{D}(\rho) \delta(1-u) - \Delta \tilde{f}_q^{MI}(u) \delta(\rho-a)\right] 
 \right. \right.\nonumber\\
& +& \hspace*{-3mm}  \left. \left.  g\left(\frac{x}{u},Q^2\right)\, \Delta D_{q_i}\left(\frac{z}{\rho},Q^2\right)\,
\frac{\alpha_s}{2\pi} T_f\left[ 
  \Phi\Delta_{gq}(u,\rho) - \Delta
\tilde{f}_g^{MI}(u) \delta (\rho-a) \right] 
\right\}  \right. \nonumber\\ 
& & \hspace*{12mm} \left.  + \, \int_{B} \frac{du}{u} (1-x)
\left\{ M\Delta_{q_i}\left(\frac{x}{u},(1-x) z,Q^2\right) \delta(1-u) \right.\right. \nonumber \\
& + & \hspace*{-3mm}\left. \left.  M\Delta_{q_i}\left(\frac{x}{u},(1-x) z,Q^2\right) \, \frac{\alpha_s}{2\pi} C_f \left[
  \Phi\Delta_{q}(u) - \Delta\tilde{f}_q^{MH}(u) \right]
 \right.\right. \nonumber \\ 
&+&\hspace*{-3mm} \left.\left. M\Delta_g \left(\frac{x}{u},(1-x) z,Q^2\right) \,
 \frac{\alpha_s}{2\pi} T_f \left[  \Phi\Delta_g(u)
-\Delta\tilde{f}_g^{MH}(u) \right] 
\right\}\right\}, 
\end{eqnarray}
which is free of collinear divergences.\\

\noindent{\large \bf 3b. Double (initial- and final-state) Polarization}\\

An analysis similar to that of subsection 3.a shields for $\Delta \sigma \Delta$ the following convolutions 
\begin{eqnarray}
\label{conv2}
\Delta \sigma \Delta \, \propto \, \, \Delta f_{j/A}(x/u)\otimes \Delta H\Delta(u,\rho) \otimes
\Delta D_{h/i_{\alpha}}(z/\rho)  \nonumber \\ 
+(1-x) \, \, \Delta M\Delta_{j,h/A} (x/u, (1-x) z) \otimes 
H' (u),
\end{eqnarray}
where again the first term corresponds to current fragmentation processes and
the second to fragmentation of the target.
In the former, $\Delta f_{j/A}(x/u)$ are now polarized parton distributions, 
$\Delta D_{h/i_{\alpha}}(z/\rho)$ are polarized fragmentation functions, and 
$\Delta H\Delta = \sum_{\lambda_1,\lambda_2} \lambda_1 \lambda_2 \, H(\lambda_1,\lambda_2)$ is
the partonic tensor for polarized initial and final states. In the second term of 
eq. (\ref{conv2}) we define the doubly polarized fracture functions as 
\begin{equation}
\Delta M\Delta   = M_{j,h/A} \left( +,+\right)-
M_{j,h/A} \left( -,-\right) - M_{j,h/A} \left( +,-\right) + M_{j,h/A} \left( -,+\right),
\end{equation}
and  tensor $ H'= \sum_{\lambda_1,\lambda_2} H(\lambda_1,\lambda_2)$,
also integrated in $\rho$.

Using now the metric and longitudinal projectors
\begin{eqnarray}
P^{\mu\nu}_M = - g^{\mu\nu} \,\,\,\,\, {\rm and} \,\,\,\,\,\,  P^{\mu\nu}_L = \frac{4 x^2}{Q^2} P^\mu P^\nu
\end{eqnarray}
and defining the metric and longitudinal components of the total cross section 
\begin{equation}
\Delta\sigma\Delta=\Delta\sigma^M\Delta+\Delta\sigma^L\Delta ,
\end{equation}
we obtain
\begin{eqnarray}
& & \hspace*{-5mm}\frac{d \Delta\sigma^M \Delta }{dx\,dy\,dz}  = 
  Y^M   \sum_{i=q,\bar q} 2\, c_i \left\{ \int \int_{A} \frac{du}{u} \frac{d\rho}{\rho}
\left\{ \Delta q_i\left(\frac{x}{u}\right) \, \Delta D_{q_i}\left(\frac{z}{\rho}\right)\,
\delta(1-u)\delta(1-\rho)\frac{}{}
\right.\right.  \nonumber \\
&+& \left.\left.
\Delta q_i\left(\frac{x}{u}\right)\, \Delta D_{q_i}\left(\frac{z}{\rho}\right) \frac{\alpha_s}{2\pi}\,
\right.\right.  \nonumber \\
\times&&  \hspace*{-12mm}\left.\left. 
 \left[ \pe  \left(\Delta P_{q\leftarrow q}(\rho)\delta(1-u)+
\Delta P_{q\leftarrow q}(u)\delta(1-\rho)\right) +C_f 
\Delta\Phi^M\Delta_{qq}(u,\rho) \right] \right.\right. \nonumber\\
&+& \left.   \left.  \Delta q_i\left(\frac{x}{u}\right)\,\Delta D_g\left(\frac{z}{\rho}\right) \frac{\alpha_s}{2\pi}\,
\right.\right.  \nonumber \\
\times& &\hspace*{-12mm} \left.\left.  
 \left[ \pe  \left(\Delta P_{g\leftarrow q}(\rho)\delta(1-u)+
   \Delta \hat P\Delta_{gq\leftarrow q}(u)\delta(\rho-a)\right) +C_f 
\Delta \Phi^M\Delta_{qg}(u,\rho)\right] 
 \right. \right.\nonumber\\
&+&  \left.  \left.  \Delta g\left(\frac{x}{u}\right)\,\Delta D_{q_i}\left(\frac{z}{\rho}\right) \frac{\alpha_s}{2\pi}\,
\right.\right.  \nonumber \\
\times &&  \hspace*{-12mm}\left.\left.
 \left[ \pe  \left( \Delta P_{q\leftarrow g}(u)\delta(1-\rho)+
   \Delta \hat P\Delta_{q\bar{q}\leftarrow g}(u)\delta(\rho-a)\right) + T_f 
\Delta \Phi^M\Delta_{gq}(u,\rho)\right] \right\}\right.  \nonumber\\
&& \hspace*{35mm}\left. +\, \int_{B} \frac{du}{u} (1-x)
\left\{  \Delta M\Delta_{q_i}\left(\frac{x}{u},(1-x) z\right)  \delta(1-u) \right. \right.\nonumber \\
&+& \left.\left.  \Delta M\Delta_{q_i}\left(\frac{x}{u},(1-x) z\right) \frac{\alpha_s}{2\pi} \left[ \pme   P_{q\leftarrow 
q}(u)+C_f  
 \right]  \right. \right.\nonumber \\
&+& \left.\left.  \Delta M\Delta_g \left(\frac{x}{u},(1-x) z\right) \,
 \frac{\alpha_s}{2\pi} \left[ \pme  P_{q\leftarrow g}(u)+ T_f 
\Delta\Phi^M\Delta_g (u)\right] \right\}\right\} \hspace*{6mm}\mbox{(36)}\nonumber
\end{eqnarray} 
\addtocounter{equation}{1}
and
\begin{eqnarray}
  \frac{d \Delta\sigma^L \Delta }{dx\,dy\,dz}  &=& 
  Y^L  \sum_{i=q,\bar q} 2\, c_i \left\{ \int \int_{A} \frac{du}{u} \frac{d\rho}{\rho}
\left\{  
  \Delta q_i\left(\frac{x}{u}\right)\, \Delta D_{q_i}\left(\frac{z}{\rho}\right)\,
\frac{\alpha_s}{2\pi}  C_f 
\Delta\Phi^L\Delta_{qq}(u,\rho) \right.\right. \nonumber\\
&+& \left.   \left.  \Delta q_i\left(\frac{x}{u}\right)\,\Delta D_g\left(\frac{z}{\rho}\right)\,
\frac{\alpha_s}{2\pi}  C_f 
\Delta \Phi^L\Delta_{qg}^A(u,\rho)
 \right. \right.\nonumber\\
&+&  \left.  \left.  \Delta g\left(\frac{x}{u}\right)\,\Delta D_{q_i}\left(\frac{z}{\rho}\right)\,
\frac{\alpha_s}{2\pi}   T_f 
\Delta \Phi^L\Delta_{gq}(u,\rho) \right\}\right.  \nonumber\\
&+& \left.  \int_{B} \frac{du}{u} (1-x)
\left\{  \Delta M\Delta_{q_i}\left(\frac{x}{u},(1-x) z\right)    \frac{\alpha_s}{2\pi}  C_f  \Delta\Phi^L\Delta_{q}(u)   \right. \right.\nonumber \\
&+& \left.\left.  \Delta M\Delta_g \left(\frac{x}{u},(1-x) z\right) \,
 \frac{\alpha_s}{2\pi}   T_f 
\Delta\Phi^L\Delta_g (u)  \right\}\right\}\hspace*{32mm} \mbox{(37)}\nonumber  
\end{eqnarray} 
\addtocounter{equation}{1}
While the longitudinal component, which is zero at leading order, is finite, the collinearly divergent metric component has yet to be subtracted, using the method indicated in the last subsection.  We then use the NLO definition of polarized parton distributions \cite{gordon,prd} and polarized fragmentation functions in the $\overline{MS_p}$ scheme, and introduce the NLO doubly polarized fracture functions by means of  
 \begin{eqnarray}
\label{2p}
  \Delta M\Delta_{q_i} (\xi,\zeta) = 
 \int_{\frac{\xi}{1-\zeta}}^1 \frac{du}{u} \left\{ 
\left[ \delta(1-u) + \frac{\alpha_s}{2\pi}  
\ppe  P_{q\leftarrow q}(u)  \right]  \Delta M\Delta_{q_i}\left(\frac{\xi}{u},\zeta,Q^2\right) \right. \nonumber \\
\left. + \frac{\alpha_s}{2\pi}  
\ppe  P_{q\leftarrow g}(u)  \Delta M\Delta_g\left(\frac{\xi}{u},\zeta,Q^2\right) \right\}
\hspace*{38mm}
\nonumber \\
 \hspace*{5mm} +\int_{\xi}^{\frac{\xi}{\xi+\zeta}} \frac{du}{u} \frac{u}{x(1-u)}
\left\{   \frac{\alpha_s}{2\pi} 
\ppme  \Delta \hat P\Delta_{gq\leftarrow q}(u)    \Delta q_i\left(\frac{\xi}{u},Q^2\right)  \,
\Delta D_g\left(\frac{\zeta u}{\xi (1-u)},Q^2\right) \right. \nonumber \\
 \left. + \frac{\alpha_s}{2\pi}  
\ppme \Delta \hat P\Delta_{q\bar{q}\leftarrow g}(u)    \Delta g\left(\frac{\xi}{u},Q^2\right) \,
\Delta D_{q_i}\left(\frac{\zeta u}{\xi (1-u)},Q^2\right) \right\}. \hspace*{4mm} \mbox{(38)}\nonumber  
\end{eqnarray}  
\addtocounter{equation}{1}
In this case no, finite terms are subtracted since the corrections for the 
fracture functions have exactly the same structure as those in the unpolarized 
case. In fact, as  can be seen in the contribution from the box diagram, 
finite soft terms (which for the box diagram are just the hat terms) coming 
from the effects of the initial- and final-state polarization in the HVBM 
prescription cancel leaving no soft contributions 
proportional to $\delta(\rho-a)$.

We then obtain the following finite metric component of the cross section
\begin{eqnarray}
\frac{d\Delta \sigma^M\Delta}{dx\,dy\,dz} & = & \nonumber \\
&& \hspace*{-9mm} Y^M   \sum_{i=q,\bar q}  2\, c_i\left\{\int \int_{A} \frac{du}{u}
\frac{d\rho}{\rho} 
\left\{ 
 \Delta q_i\left(\frac{x}{u},Q^2\right)\, \Delta D_{q_i}\left(\frac{z}{\rho},Q^2\right) \,
\delta(1-u)\delta(1-\rho)\frac{}{} 
\right. \right.  \nonumber \\
&+& \left.\left.
\Delta q_i\left(\frac{x}{u},Q^2\right)\, \Delta D_{q_i}\left(\frac{z}{\rho},Q^2\right)  \,
\right. \right.  \nonumber \\
&&  \hspace*{-2mm}\left.\left. \times \,
\frac{\alpha_s}{2\pi} C_f \left[ 
   \Delta\Phi^M\Delta_{qq}(u,\rho) - \Delta \tilde{f}_q^D(\rho) \delta(1-u) - \Delta \tilde{f}_q^F(u) \delta(1-\rho) \right] \right.\right.
\nonumber \\   
&+&  \left.\left.  \Delta q_i\left(\frac{x}{u},Q^2\right)\, \Delta D_g\left(\frac{z}{\rho},Q^2\right)\,
\right. \right.  \nonumber \\
&&  \hspace*{-2mm}\left.\left. \times \,
\frac{\alpha_s}{2\pi} C_f \left[
 \Delta\Phi^M\Delta_{qg}(u,\rho)- \Delta \tilde{f}_g^{D}(\rho) \delta(1-u) \right] 
 \right. \right.\nonumber\\
& +&   \left. \left. \Delta g\left(\frac{x}{u},Q^2\right)\, \Delta D_{q_i}\left(\frac{z}{\rho},Q^2\right)\,
\right. \right.  \nonumber \\
&&  \hspace*{-2mm}\left.\left. \times \,
\frac{\alpha_s}{2\pi} T_f\left[ 
  \Delta\Phi^M\Delta_{gq}(u,\rho) - \Delta
\tilde{f}_g^{F}(u) \delta(1-\rho)  \right] 
\right\}  \right. \nonumber\\ 
& &\hspace*{20mm}\left. + \,  \int_{B} \frac{du}{u} (1-x)
\left\{ \Delta M\Delta_{q_i}\left(\frac{x}{u},(1-x) z,Q^2\right) \delta(1-u)  \right. \right. \nonumber \\
& +& \left. \left.  \Delta M\Delta_{q_i}\left(\frac{x}{u},(1-x) z,Q^2\right)\frac{\alpha_s}{2\pi} C_f  
 \Delta \Phi^M\Delta_{q}(u)  \right.\right. \nonumber \\ 
&+ & \left.\left. \Delta M\Delta_g \left(\frac{x}{u},(1-x) z,Q^2\right) \,
 \frac{\alpha_s}{2\pi} T_f  \Delta \Phi^M\Delta_g(u) 
\right\}\right\} \hspace*{23mm} \mbox{(39)} \nonumber
\end{eqnarray}
\addtocounter{equation}{1}
where the finite terms subtracted from the polarized parton distributions are
\begin{eqnarray}
   \Delta\tilde{f}_q^{F}(u) = 4\,(u-1)  \,\,\,\, {\rm and} \,\,\,\,
  \Delta\tilde{f}_g^{F}(u)  = 2\,(1-u) 
 \end{eqnarray}
and the coefficients can be found in the Appendix.

Notice that both the single polarized (final-state) fracture function
$M \Delta$ and the doubly polarized one $\Delta M \Delta$ are not mere artefacts motivated by factorization, but they are physically motivated 
parton densities related to target fragmentation. 
In the latter case they 
parametrize unpolarized parton densities of a nucleon that fragments into
a hadron with a definite polarization state. In the former case, they 
involve polarized parton distributions of fragmented nucleons\footnote{The notation used may be, up to a point, misleading: ``double'' and ``single'' refer to
a given combination of cross sections that will be proportional, in the end, to one or the
other fracture function. The partonic pictures for fracture functions  are given by eqs. (25) and (33), also confirmed  by the  kernels that drive
their evolution (polarized or unpolarized) in eqs. (27) and (36).   }. 

At this stage, it is worth noticing that any factorization scheme can be chosen as long as the  distributions used  had been obtained by means of a consistent NLO analysis (with the corresponding
splitting functions and coefficients in the same scheme). For instance, in the case of polarized parton distributions, NLO parametrizations can be obtained from ref. \cite{forteball} in the $\overline{MS_p}$ scheme, or from refs. \cite{gs,grsv}, in a mixed factorization scheme where only the $\Delta \tilde{f}^F_q(u)=4 (u-1)$ has been substracted ($\Delta \tilde{f}^F_g(u)=0$ in eq. (40)). In the case of polarized fragmentation functions, no parametrizations are available yet, but the same  concepts apply to them, both in the case of coputations using dimensional regularization or off-shell gluons, as in ref. \cite{rav}, to regularize the divergencies. The corresponding NLO polarized splitting functions have been obtained very recently \cite{vogstrat} in the mixed factorization scheme ($\Delta \tilde{f}^D_q(y)=4 (y-1)$ and $\Delta \tilde{f}^D_g(u)=0$ in eq. (14)) and will allow for a complete NLO analysis of polarized fragmentation functions.

\noindent{\large \bf 4. Summary and Conclusions}\\

Summarizing our results, we have presented here the full $\cal{O}$$(\alpha_s)$ 
QCD corrections to the semi-inclusive production cross section of polarized 
hadrons in $e^+e^-$ annihilation processes and polarized deep inelastic
lepton-hadron scattering using dimensional regularization and the HVBM 
scheme. 

Generalizing the factorization prescription that excludes soft
contributions in polarized deep inelastic ($\overline{MS_p}$) to polarized 
fragmentation functions and to two new types of fracture functions, we have 
shown that this prescription smoothly subtracts all the collinear divergences 
and soft contributions present in the semi-inclusive production of polarized
hadrons in polarized DIS without introducing kinematical
cuts or other process-dependent procedures. We present  for 
completeness also  our results in the more familiar $\overline{MS}$ scheme, that offer
the same advantages with respect to factorization, but with a less trivial
interpretation for parton distributions.

In the case of polarized fragmentation functions, we find our results compatible with the ones presented in ref. \cite{rav}, computed using a different regularization prescription. Of course, both results can be consistently used as long as the scheme dependence is cancelled  by the use of the corresponding NLO splitting functions in the chosen scheme. 

The evolution kernels obtained here for fracture functions also allow  to study
the pertubative component of the scale dependence, at least to leading order, 
of these densities, unveiling new aspects of target fragmentation processes
induced by DIS. 
In this way we produce a fully consistent, physically motivated and complete 
set of QCD-corrected cross sections, of interest in future phenomenological 
approaches and forthcoming experiments.\\

\newpage

\noindent{ \bf Acknowledgements}

We gratefully acknowledge C. Garc\'{\i}a Canal, G. Veneziano and 
D. Graudenz  for enlightening comments and discussions. \\

\noindent{\large \bf  Appendix }\\

We append here the coefficients, some definitions and useful relations.

a. \underline{Coefficients for single polarization}
\begin{eqnarray}
 && \Phi\Delta_{qq}(u,\rho) =   \nonumber \\
&-&  8\,  \delta (1-u) \delta (1-\rho)  
   -  \frac{1}{(1-u)_+} (1+\rho)
   -  \frac{1}{(1-\rho)_+} (1+u) \nonumber \\
&+&    \delta (1-u) \left[ \rho-1-(1+\rho) \ln (1-\rho) + \frac{1+\rho^2}{1-\rho} \ln\rho + 2 \left( \frac{\ln (1-\rho)}{1-\rho}\right)_+ \right] \nonumber \\
&+&    \delta (1-\rho) \left[ 1 - u -(1+u) \ln (1-u) + \frac{1+u^2}{1-u}\ln\frac{1-x}{u-x} + 2 \left( \frac{\ln (1-u)}{1-u}\right)_+ \right] \nonumber \\
&+&    2 \frac{1}{(1-\rho)_+} \frac{1}{(1-u)_+} +   2\,\left( 2 + u \right) \,\left( 1 - x \right)  - 
     2\,\left( 2 - \rho \right) \,{{\left( 1 - x \right) }^2}\nonumber \\
&+&    
     {{x\,\left( 6\, x\, (1-\rho) (1-x) +x\, (1-\rho)-2+2\,\rho\, (1-x^2)\right) } \over {u - x}} \nonumber \\
&+&    {{\left( 1 - \rho \right) \,\left( 1 - x \right) \,{x^2}\,
         \left(  2\,x -1  \right) }\over {{{\left( u - x \right) }^2}}}   
\end{eqnarray}
\begin{eqnarray}
\Phi\Delta_{qg}(u,\rho) &=&  \delta (1-u) (2-\rho)\ln[(1-\rho)\,\rho] +   \left(\frac{1}{1-u}\right)_+ ( 2 -\rho)
    \nonumber \\
 &+&   \delta (\rho-a) (1+u)  \ln\frac{1-u}{u} +  \left(\frac{1}{\rho-a}\right)_+ (1+ u) \nonumber \\
&-&  2 (1-\rho)^2 (1+\rho^2)\, \widehat{\delta (1-u)}    -1 + 
     {{2\,\left( 1 -\rho \right) \,\left( 2\,x - 1 \right) }\over {1 - x}} \nonumber \\
&+& {{u -2 \, u^2 (1-x) }\over {u - x}} -  {{ 2 \left( 1 - \rho \right) \,\left( 1 - u \right) \, \, x \, u \, (1-x) }\over {  \,{{\left( u - x \right) }^2}}} 
   \nonumber \\
&+&       {{ 2\left( 1 - \rho \right) \,\left( 1 - u \right)  \,x^2\,
         }\over {\left( 1 - x \right) \,{{\left( u - x \right) }}}} +   {{\left( 1 - \rho \right) \,\left( 1 - u \right) \,x^2\,
           }\over {  \,{{\left( u - x \right) }^2}}}  -  2  \widehat{\delta (\rho-a)} \nonumber \\
\end{eqnarray}
\begin{eqnarray}
\Phi\Delta_{gq}(u,\rho) &=&   
\delta(1-\rho)\left[1+(1-2u+2u^2)\left(-1+\ln\frac{(1-u)(1-x)}{u-x}\right)
\right] \nonumber \\
&+&  \left(\frac{1}{1-\rho}\right)_+ (1-2 u+2u^2 )
  +  2(1-u) \widehat{\delta (\rho-a)} \nonumber  \\
 &+&   \delta (\rho-a)
\left[1-(1-2u+2u^2)\left(-1+\ln\frac{1-u}{u}\right)\right]
 \nonumber \\ 
&-&  \left(\frac{1}{\rho-a}\right)_+ (1-2u+2u^2) 
\end{eqnarray}

\begin{eqnarray}
 \Phi\Delta_{q}(u) & = & 
-\frac{1+u^2}{1-u}\ln(u)-(1+u)\ln(1-u)
+2\left(\frac{\ln(1-u)}{(1-u)} \right)_{+}  \nonumber \\
&-& \frac{3}{2} \left(\frac{1}{1-u}\right)_+ + 3 u + (-\frac{9}{2}-\frac{\pi^2}{3})\delta(1-u) -2\widehat{(1-u)}
\end{eqnarray}
\begin{eqnarray}
 \Phi\Delta_{g} (u) &=& 2\widehat{(1-u)} +(2 u-1) 
\left[ \ln \left( {{1-u}\over u} \right) -1    \right]  
\end{eqnarray}

b. \underline{Coefficients for double  polarization}

i. \underline{Metric}
\begin{eqnarray}
&& \Delta\Phi^M\Delta_{qq}(u,\rho)=   \nonumber \\ 
&-&  8\,  \delta (1-u) \delta (1-\rho)    -  \frac{1}{(1-u)_+} (1+\rho) 
   -  \frac{1}{(1-\rho)_+} (1+u) \nonumber \\
&+&    \delta (1-u) \left[ \rho-1-(1+\rho) \ln (1-\rho) + \frac{1+\rho^2}{1-\rho} \ln\rho + 2 \left( \frac{\ln (1-\rho)}{1-\rho}\right)_+ \right] \nonumber \\
&+&   \delta (1-\rho) \left[ u - 1 -(1+u) \ln (1-u) + \frac{1+u^2}{1-u}\ln\frac{1-x}{u-x} + 2 \left( \frac{\ln (1-u)}{1-u}\right)_+ \right] \nonumber \\
&+&   2 \frac{1}{(1-\rho)_+} \frac{1}{(1-u)_+} +   {{\left( 1 - \rho \right) \,x\,\left( 2\,u - x - u\,x \right) }\over {{{\left( u - x \right) }^2}}}\nonumber \\
 &+&  {{2\,\left( u - x - u\,x \right) }\over {u - x}} 
 \end{eqnarray}
\begin{eqnarray}
\Delta \Phi^M\Delta_{qg}(u,\rho) &=&  \delta (1-u) (2-\rho)\ln[(1-\rho)\, \rho] 
 +   \left(\frac{1}{1-u}\right)_+ ( 2 -\rho)
  \nonumber \\
 &+&    \delta (\rho-a) (1+u)\left[2+\ln\frac{1-u}{u}\right] +  \left(\frac{1}{\rho-a}\right)_+ (1+ u)  \nonumber \\
&-& 2\,  (1-\rho^2)\, \widehat{\delta (1-u)} 
   -2\,\left( 1 - x \right)  + {{\left( 1 - \rho \right) \,x}\over {u - x}} - {{\rho\,\left( 1 - x \right) \,x}\over {u - x}} \nonumber \\
&+&   {{\left( 1 - \rho \right) \,\left( 1 - x \right) \,{x^2}}\over 
       {{{\left( u - x \right) }^2}}} + {{{x^2}}\over {u - x}} -  2\, u \, \widehat{\delta (\rho-a)}    
\end{eqnarray}
\begin{eqnarray}
\Delta \Phi^M\Delta_{gq}(u,\rho) &=&  \delta (1-\rho) (2u-1)\left(\ln\frac{(1-u)(1-x)}{u-x}\right)  \nonumber \\
 &+&   \left(\frac{1}{1-\rho}\right)_+ (2u-1 )-  \left(\frac{1}{\rho-a}\right)_+ (2u-1)
  \nonumber \\ 
&+&     \delta (\rho-a)
(2u-1)\left(-2-\ln\frac{1-u}{u}\right) \nonumber \\
&+& 2(1-u) \widehat{\delta (1-\rho)} 
   \end{eqnarray}
\begin{eqnarray}
\Delta\Phi^M\Delta_{q}(u) & = & 
-\frac{1+u^2}{1-u}\ln(u)-(1+u)\ln(1-u)
+2\left(\frac{\ln(1-u)}{1-u} \right)_{+} \nonumber \\ 
&-&\frac{3}{2}
 \left(\frac{1}{1-u}\right)_+  + 3 -u + (-\frac{9}{2}-\frac{\pi^2}{3})\delta(1-u)  \end{eqnarray}
\begin{eqnarray}
\Delta\Phi^M\Delta_g (u)= (1-2 u+2u^2) \left(\ln\frac{1-u}{u} -1  \right)
\end{eqnarray}

  ii. \underline{Longitudinal} 
\begin{eqnarray}
\Delta\Phi^L\Delta_{qq}(u,\rho)=\frac{2 (\rho-a) u^3 (1-x)^3}{(u-x)^2}
\end{eqnarray}
\begin{eqnarray}
\Delta \Phi^L\Delta_{qg}(u,\rho)=-2(1-\rho) u^3 \frac{(1-x)^2 }{(u-x)^2}  
\end{eqnarray}
\begin{eqnarray}
\Delta \Phi^L\Delta_{gq}(u,\rho)= 0
\end{eqnarray}
\begin{eqnarray}
\Delta\Phi^L\Delta_{q}(u)= u
\end{eqnarray}
\begin{eqnarray}
\Delta\Phi^L\Delta_g (u)= 2u(1-u)  
\end{eqnarray}

c. \underline{Splitting functions}

We include here the expressions for the virtual \cite{ap} and real \cite{uemven} splitting functions:
 \begin{eqnarray}
\Delta P_{q\leftarrow q}(u)&=&   P_{q\leftarrow q}(u)=
C_f \left[ 2\left(\frac{1}{1-u}\right)_+ + \frac{3}{2}
\delta(1-u)-1-u\right]
\nonumber \\
\Delta P_{q\leftarrow g}(u)&=& \Delta \hat P_{q\bar{q}\leftarrow g}(u) = - \Delta \hat P\Delta_{q\bar{q}\leftarrow g}(u) = T_f \left[2u-1\right]\nonumber \\
 P_{q\leftarrow g}(u)&=&   - \hat P\Delta_{q\bar{q}\leftarrow g}(u) = T_f \left[2u^2 -2 u+1\right] \\
 \Delta P_{g\leftarrow q}&=& \hat P\Delta_{gq\leftarrow q}(1-u) = \Delta \hat P\Delta_{gq\leftarrow q}(1-u) =
C_f \left[ 2-u\right]\nonumber \\
 P_{g\leftarrow q}(u)&=& \Delta\hat P_{gq \leftarrow q}(1-u) = C_f \left[ 2\frac{1}{u} -2+u \right] \nonumber
  \end{eqnarray}

d. \underline{Coefficients for single initial-state polarization}

We include here the coefficients from ref. \cite{npb} adapted to 
the definition of the $()_+$ prescriptions used in ref. \cite{graudenz}:
\begin{eqnarray}
& & \Delta\Phi_{qq}(u,\rho) = \nonumber \\
& - & 8\, \delta(1-\rho) \delta(1-u)  -
\left(\frac{1}{1-\rho}\right)_+ (1+u)- \left(\frac{1}{1-u}\right)_+(1+\rho)
\nonumber \\ 
&+& \left[(1-\rho)+\frac{1+\rho^2}{1-\rho}
\ln\rho-(1+\rho)\ln(1-\rho)+2\left(\frac{\ln(1-\rho)}{1-\rho} \right)_{+}\right]
\delta(1-u)\nonumber\\
&+& \left[-(1-u)+\frac{1+u^2}{1-u}
\ln(\frac{1-x}{u-x})-(1+u)\ln(1-u)+2\left(\frac{\ln(1-u)}{1-u} \right)_{+}
\right]
\delta(1-\rho) \nonumber\\
&-&{{\left( 1 - \rho \right) \,\left( 1 - u + 
        \left( 1 - x \right) \,\left( -1 + 
           u\,\left( 1 + 2\,u \right) \,\left( 1 - x \right)  - x \right) 
 \right) }\over {{{\left( u - x \right) }^2}}} + 
  {{4\,u\,\left( 1 - x \right) }\over {u - x}} \nonumber \\
&-&   {{2\,\left( 1 - u \right) \,u\,\left( 1 - x \right) }\over {u - x}}-  {{2\,x}\over {u - x}} + 2 \left(\frac{1}{1-\rho}\right)_+ \left(\frac{1}{1-u}\right)_+ -2 (1-u) \widehat{\delta(1-\rho)}\nonumber \\
\end{eqnarray}
\begin{eqnarray}
\Delta \Phi_{qg}(u,\rho) & = & \delta (1-u) \left[\rho +  \left(\rho+{2\over \rho}-2 \right)
\ln \left(\rho(1-\rho) \right) \right] \nonumber
\\
&+& \delta (\rho-a) \left[-2\widehat{(1-u)}-(1-u) +
{{1+u^2}\over{1-u}} \ln \frac{1-u}{u} \right]
- {{2\,{{\left( 1 - \rho \right) }^2}}\over {\rho\,\left( 1 - u \right) }}
\nonumber \\
 &+& 
  {{ -2\, u}\over { (\rho-a)(1-u)}} - {{2\,\left( 1 - u \right) \,u\,
      \left( 1 - x \right) }\over {u - x}} + 
  {{2\,{{\left( 1 - \rho \right) }^2}\,{u^3}\,{{\left( 1 - x \right) }^2}}\over
      {\left(  \rho-a \right) \,\left( 1 - u \right) \,
      {{\left( u - x \right) }^2}}}  \nonumber\\
& +&   {{\left( 2 - \rho \right) \,x}\over {u - x}}
+\left({1\over{\rho-a}}\right)_+  {{1+u^2}\over{1-u}} +\left({1\over{1-u}}\right)_+ \left(\rho+{2\over \rho}-2 \right)  \nonumber \\
&+& {{\left( 1 - \rho \right) \,u\,\left( 1 - x \right) \,x}\over 
    {{{\left( u - x \right) }^2}}}  
\end{eqnarray}
\begin{eqnarray}
\Delta\Phi_{gq}(u) &=& \delta (1-\rho) \left[2\widehat{(1-u)} + (2 u-1) \ln \left(
{{(1-x)(1-u)} \over {u-x }} \right) \right]  \nonumber  \\
&+& \delta (\rho - a) \left[2\widehat{(1-u)} + (2 u-1) \ln\frac{1-u}{u} \right]  \nonumber \\
& + & (2 u-1) \left[ \left({1\over{1-\rho}}\right)_+
+\left({1\over{\rho-a}}\right)_+ \right] -2 u{{1-x}\over{u-x}}  
\end{eqnarray}
\begin{eqnarray}
\Delta \Phi_{q}(u) & = & 
-\frac{1+u^2}{1-u}\ln(u)-(1+u)\ln(1-u)
+2\left(\frac{\ln(1-u)}{1-u} \right)_{+} \nonumber \\ 
&-&\frac{3}{2}
 \left(\frac{1}{1-u}\right)_+ 
 + 3 u + \left(-\frac{9}{2}-\frac{\pi^2}{3}\right)\delta(1-u) -2\widehat{(1-u)}
\end{eqnarray}
\begin{eqnarray}
\Delta \Phi_{g} (u,\rho)& = & 2\widehat{(1-u)} +(2 u-1) 
\left[ \ln \left( {{1-u}\over u} \right) -1  \right]  
\end{eqnarray}

\newpage

\newpage
 \begin{figure}[t]
\begin{center}
\mbox{\kern-2cm
\epsfig{file=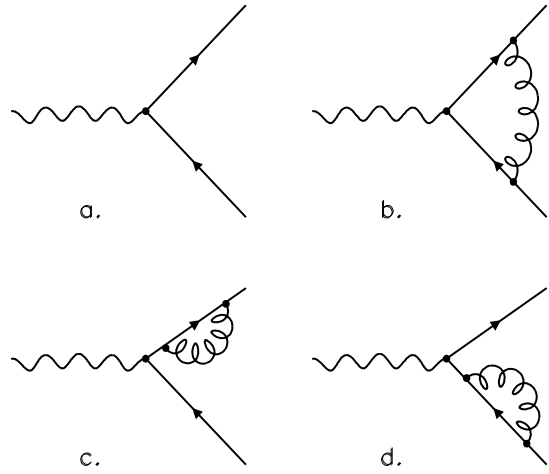,width=10.0truecm,angle=0}}
\vspace{-2cm}
\caption{{\bf a}: Lowest order diagram for $e^+ e^-$ anihilation; {\bf b, c, d}: Virtual gluon corrections to {\bf 1a }
 }
\end{center}
\end{figure}
 \begin{figure}[t]
\begin{center}
\mbox{\kern-2cm
\epsfig{file=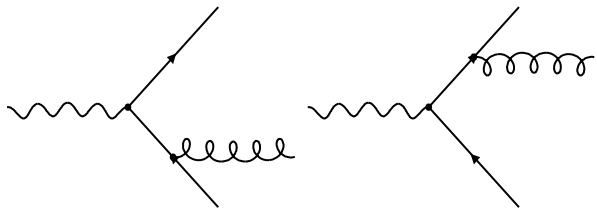,width=10.0truecm,angle=0}}
\vspace{-5cm}
\caption{Real gluon emission corrections to $e^+ e^-$ anihilation  }
\end{center}
\end{figure}
\newpage
 \begin{figure}[t]
\begin{center}
\mbox{\kern-2cm
\epsfig{file=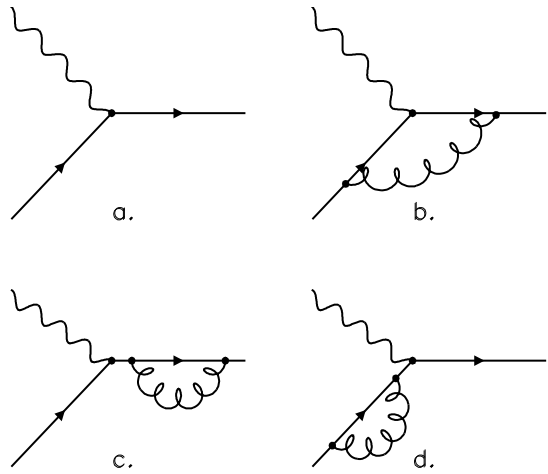,width=10.0truecm,angle=0}}
\vspace{-2cm}
\caption{ {\bf a}: Lowest order parton-photon graph; {\bf b}, {\bf c}, {\bf d}: Virtual gluon corrections graphs to {\bf 3a}}
\end{center}
\end{figure}
 \begin{figure}[t]
\begin{center}
\mbox{\kern-2cm
\epsfig{file=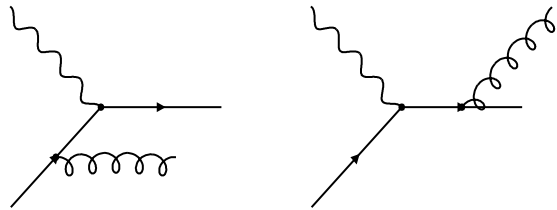,width=10.0truecm,angle=0}}
\vspace{-5cm}
\caption{Real gluon emission corrections to DIS
 }
\end{center}
\end{figure}
\newpage
 \begin{figure}[t]
\begin{center}
\mbox{\kern-2cm
\epsfig{file=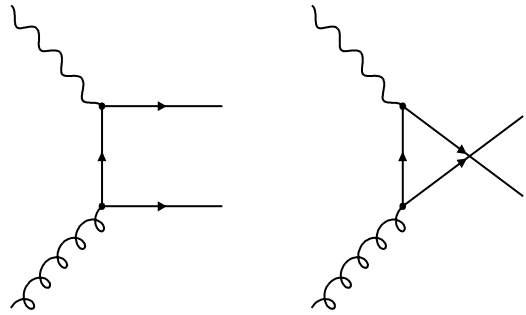,width=10.0truecm,angle=0}}
\vspace{-2cm}
\caption{ Gluon contribution (box diagram) to DIS}
\end{center}
\end{figure}

\end{document}